# Spectroscopic investigation of defects in two dimensional materials

Zhangting Wu and Zhenhua Ni[*]

State Key Laboratory of Bioelectronics, Department of Physics, Southeast University, Nanjing 211189, China.

**Abstract:**

Two-dimensional (2D) materials have been extensively studied in recent years due to their unique properties and great potential for applications. Different types of structural defects could present in 2D materials and have strong influence on their properties. Optical spectroscopic techniques, e.g. Raman and photoluminescence (PL) spectroscopy, have been widely used for defect characterization in 2D materials. In this review, we briefly introduce different types of defects and discuss their effects on the mechanical, electrical, optical, thermal, and magnetic properties of 2D materials. Then, we review the recent progress on Raman and PL spectroscopic investigation of defects in 2D materials, i.e. identifying of the nature of defects and also quantifying the numbers of defects. Finally, we highlight perspectives on defect characterization and engineering in 2D materials.

* Corresponding author: zhni@seu.edu.cn

## 1. Introduction

Two-dimensional (2D) materials have been extensively studied in the past decade due to their unique properties. Graphene as a typical 2D material has already shown great potential in mechanics, thermotics and electrics: it is the thinnest known material and the strongest ever measured [1]; it shows superior thermal conductivity, i.e. 5000 $Wm^{-1}K^{-1}$ [2]; its charge carriers approach ballistic transport and can travel for micrometers without scattering [3]. The research progress in graphene has also triggered enthusiasm to investigate many other 2D materials, including 2D transition metal dichalcogenides (TMDs), boron nitride, black phosphorus and so on [4-7]. Especially, 2D TMDs, e.g. $MoS_2$, have attracted great attentions and are considered as promising candidates for next-generation electronics and optoelectronics [8-10]. They are semiconductor materials with tunable bandgap in the range of 1-2 eV, and undergoing a transition from indirect to direct band gap with reduced thickness to monolayer. Both high on/off ratio in logic transistor and high gain of photoresponse have been achieved for monolayer or multilayer TMDs semiconductor [11, 12].

Like all other materials, the structures of 2D materials always contain abundant and different types of defects, such as vacancies, adatoms, edges, grain boundaries (GBs), and substitutional impurities, which would strongly influence their properties [13]. In electrics, the carrier mobility of graphene strongly depends on the numbers of defects inside and also its crystalline grain size, due to the scattering from localized defects and GBs. It has been demonstrated that even very small amount of defects in mechanically exfoliated graphene can strongly limit its carrier mobility [14].While graphene grown by chemical vapor deposition (CVD) normally has mobility much lower than that of mechanically exfoliated sample, due to the appearance of more defects and GBs [15-19]. Similarly, the carrier mobility in TMDs devices is also limited by structural defects, which in turn restrict its electronic and optoelectronic performance [20]. Direct evidence by high resolution transmission electron microscope (TEM) has revealed that structural defects (sulphur vacancies) exist in $MoS_2$, which introduce localized donor states inside the bandgap and result in hopping transport at room temperature [21]. In mechanics, the mechanical properties of 2D materials can be affected by both the density and arrangement of defects, especially, GBs [22, 23]. The GBs in graphene can either increase or decrease the strength of graphene, depends on the detailed atomic arrangement of GBs [22, 23]. In thermotics, it has been demonstrated that the thermal conductivity of graphene could be significantly reduced even at extremely low defect concentration (~83% reduction for ~0.1% defects), which could be attributed to the creation of oxygen related defects by oxygen plasma irradiation [24].

On the other side, defects in 2D materials can also have a beneficial impact on their properties. If the location of defects can be controlled, novel graphene-based materials can be prepared for application in spintronic devices, catalyst, PN junction and so on [25-27]. By defect engineering, the bandgap of graphene can be opened to allow switching of graphene-based transistors with a high on/off ratio [13]. The properties of TMDs can also be tailored by introduction of defects, for example: line defects can act as one-dimensional metallic stripes [28]; laser and ion irradiation can be utilized to thin and dope TMDs [29,30]; GBs influence the electroluminescence (EL) behavior of $WS_2$ [31]; structural defects or active edge sites can be applied as electrocatalysis [32]; a strong photoluminescence (PL) enhancement of monolayer $MoS_2$ can be realized through defect engineering and oxygen bonding [33]. Hence, the investigation of defects is a crucial step for 2D materials research.

To thoroughly study the structural defects in 2D materials, modern characterization techniques have been employed, such as TEM, scanning tunneling microscope (STM), and X-ray photoelectron spectroscopy (XPS). Although TEM and STM provide structural image in the atomic scale, they have the problems of complicated sample preparation and small inspection areas [34-37]. While a statistic method such as XPS has a disadvantage of poor spatial resolution [38]. Optical spectroscopic methods, such as Raman and PL spectroscopies, are ideal for the characterization of 2D materials because they are time efficient and nondestructive. Raman and PL spectroscopy reveal information on the crystal structure, electronic structure and lattice vibrations, and can be used to probe thickness (or layer numbers), strain, structural stability, defects, charger transfer, and stacking orders of 2D materials [39-43]. Raman spectroscopy has been demonstrated as an efficient tool on defect characterization in graphene, due to the appearance of defect related D and D′ peaks in the spectrum of disordered sample [44]. These Raman peaks are activated by double resonant intervalley and intravalley scattering processes, respectively, where defect provides the missing momentum in order to satisfy momentum conservation in the Raman scattering process [45, 46]. For TMDs, some new Raman peaks also appear after the introduction of defects and their intensities are related to the density of defects [47], but the sensitivity is not very high. At same time, in low temperature PL measurement, a defect related PL peak of TMDs could be observed due to the emission of excitons bound to the defect sites, which could be effectively used to monitor the numbers of defects in TMDs [48].

This article will mainly review the defect characterization of 2D materials by optical spectroscopic approaches. The organization of the review is as follows: In section 2, we summarize the types of defects and discuss their effects on the properties of 2D materials; In section 3, we mainly focus on the Raman spectroscopic investigation of defects in graphene and TMDs, including the characterization of the types of defects and quantification the numbers of defects; In section 4, we will discuss the application of PL spectroscopy in the characterization of defects in TMDs. In section 5, we will draw a conclusion and give perspectives on the future studies of defects characterization and engineering in 2D materials. We hope this review will be helpful to the understanding the defects in graphene and TMDs, and also benefit future study and applications of 2D materials.

## 2 The types of defects and their effects on the properties of 2D materials.
### 2.1. The types of defects

In 2D materials, both intrinsic defects and extrinsic defects such as foreign atoms may exist. Here, defects are mainly classified according to dimensionality, i.e. zero-dimensional defects (Stone–Wales defects, vacancy, adatoms, and substitutional impurities), and one-dimensional defects (line defect, GBs and edges).

### 2.1.1. Zero-dimensional defects

**Stone–Wales (SW) defects:** in $sp^2$-hybridized hexagonal carbon systems, a simple type of defects is SW defects [49, 50], which are generated by reconstruction of graphene lattice (switching between pentagons, hexagons, and heptagons) with rotations of C-C bonds by 90° as shown in Figure 1A-i [51]. The SW defects can be formed by rapidly cooling from high temperature or under electron-beam irradiation [13, 52]. Different from graphene, the SW rotational defects are not formed in TMDs due to the polar nature of chemical bonds with trigonal symmetry. However, a 'trefoil'-shaped defect, same as $V_2$(555-777) defect in graphene, is formed by a 60° rotation of three bonds centered on a metal atom ($T_1$) in atom-deficient TMDs under e-beam irradiation and at elevated temperature, as shown in Figure 1A-ii [34]. After multiple M-X bond rotations, a larger second-order rotational defect ($T_2$) can also form [34].

**Vacancies:** especially, single vacancy (SV) is widely studied in 2D materials, as shown in Figure 1B [21]. Besides the simple SV defect, larger and more complex defect configurations appear when more than two neighboring atoms are missing. For example, double vacancies (DV) can be created either by the coalescence of two SVs or by removing two neighboring atoms in graphene. Different kinds of vacancy defects are also found in TMDs. For example, in $MoS_2$, there can be S (single S atom, $V_S$), $S_2$ (double S atoms, $V_{S2}$), or Mo vacancy ($V_{Mo}$) [28]. However, some vacancy defects are unstable, resulting in defect reconstructions in 2D materials. In graphene, SV undergoes a jahn-teller distortion, leading to the formation of a five-membered and a nine-membered ring [51, 53]. DV can be reconstructed and result in the formation of two pentagons and one octagon [$V_2$(5-8-5) defect]. In TMDs, once $V_{Mo}$ is generated, the S atoms around it become strongly prone to lose. This could be the reason $V_{Mo}$ is not observed alone, and most Mo vacancies are present as defect complexes of $V_{MoS3}$ [28].

**Adatoms:** foreign atoms could be introduced in 2D materials and form adatoms defects. If the interaction between the foreign atom and 2D materials is weak, only physisorption occurs as shown in Figure 1C-i and ii [33]. If the interaction is stronger, covalent bonding between the foreign atom and the nearest atom of 2D materials leads to chemisorption as shown in Figure 1C-iii and iv [33]. Common physisorption includes water, oxygen, polymer molecular and metal atoms on top of 2D materials, which become charge donors or acceptors and be easily removed by vacuum pumping or annealing [54-56]. Chemisorption in graphene is mainly located at three positions: in carbon-carbon bond, above carbon atoms, trapped by structural defects [57]. If atom is chemically adsorbed on graphene, $sp^3$-hybridization can be formed because of covalent bonding between foreign atom and carbon atom. Generally, the $sp^3$-hybridization in graphene could be introduced by hydrogenation, fluorination, mild oxidation, and so on [58]. The chemisorption in TMDs is more complex. In the case of 1H phase TMDs, there are four positions available for an adatom: on top of the metal or chalcogen atoms, and the hollow site slightly above the center of the hexagon on the metal or chalcogen atomic plane [59]. Studies show that chemical adsorption of oxygen molecule on S vacancy of $MoS_2$ sample is stable due to high binding energy, which can be introduced by oxygen plasma and annealing [33].

**Substitutional Impurities:** foreign atoms can also be incorporated into the lattice of 2D materials as substitutional impurities. Substitutional dopants are expected to be very stable due to strong covalent bonding as shown in Figure 1D [60-, 62]. Replacing carbon atom by transition metal impurities, boron, or nitrogen atoms can move the position of the Fermi level and change the electronic structure of graphene [61-63]. By introducing other atomic species into the TEM chamber, substitutional defects have been found in TMDs in the process of electron-beam irradiation [64]. The LDOS shows that N, P, As, and Sb atoms behave as acceptors, whereas F, Cl, Br, and I atoms are likely to be donors [64]. The isoelectronic species like O, Se, or Te atoms do not produce any localized states, which illustrates that they can heal the electronic structure of TMDs with chalcogen vacancies [64].

### 2.1.2. One-dimensional defects

**Line defects:** in graphene, two 5-7 pair defects (Figure 1E-i) with lower total energy are more stable than a local haeckelite structure (555-777) and become dislocation line defects [65]. In TMDs, S vacancies in $MoS_2$ sample are found to be mobile under electron-beam irradiation and tend to agglomerate into line defects as shown in Figure 1E-ii [66]. Single and double line vacancies are observed experimentally and both aligned along the zig-zag direction [66].

**Grain boundaries:** large-scale 2D material films synthesized by CVD are typically polycrystalline, and GBs is important type of defects that could strongly influence the properties. Figure 1F-i shows a dark-field TEM image of a graphene sheet, which illustrates that the single crystal grains have complex shapes and many different crystal orientations, with GBs clearly identified [67]. Figure 1F-ii shows two graphene grains meet with a relative misorientation of 27° and are stitched together by a series of pentagons, heptagons and distorted hexagons, forming a tilt boundary [67]. The GBs in graphene are not straight, and the defects along the boundary are not periodic.

In TMDs, various dislocation core structures constitute the GBs, including not only the topologically conventional one with five- and 7-fold (5|7) rings but also new core structures with 4|4, 4|6, 4|8, and 6|8 fold rings, which are distinctly different from GBs in graphene [68]. Two common GBs in $MoS_2$ crystals are tilt and mirror twin boundaries [69]. Figure 1F-iii and iv show dark-field TEM images of polycrystalline $MoS_2$ islands grown by CVD, with a tilt (iii) and mirror (iv) twin boundary, respectively [69]. A $MoS_2$ tilt boundary is commonly formed from line 5|7 rings, whereas twin boundary has been observed as 8|4|4 rings. In addition, boundaries composed of different mixed rings and angles can present in TMDs [68].

**Edges:** different kinds of edge terminations exist in 2D materials, which directly determine the morphologies and properties of 2D materials. As a perfect single crystalline structure, the mechanically exfoliated graphene sheets have crystal cleavage behaviors [70]. Studies found that most of the angles of graphene are distributed around $n \times 30^o$, where n is an integer between 0 and 6, which suggests that graphene has zigzag or armchair edges [70]. In TMDs, the shape of monolayer $MoS_2$ sheet evolutes from dodecagonal shape, then to hexagonal shape, and finally to triangular shape with the variation of edges from zigzag/armchair to purely zigzag

[71]. A high chemical potential of sulfur leads to triangular shape terminated by Mo-edge surface in MoS$_2$ during CVD growth, which is controlled by temperature and the ratio of partial pressures of H$_2$S and H$_2$ [72]. Similar to MoS$_2$, by controlling H$_2$ pressure to produce appropriate fluxes of WO$_{3-x}$ and Se, both hexagonal and triangular WSe$_2$ crystals can be grown [73]. However, not all edges of 2D materials are microscopically ordered, and often consist of zigzag and armchair segments [74]. Edge reconstruction was occasionally observed in TMDs. As shown in Figure 1H-i, the edge has a regular Mo-terminated structure, while in Figure 1H-ii, the outmost row of Mo atoms undergo a strong reconstruction, moving closer to the inner Mo row [28]. Generally, electronic and magnetic properties are different for different kinds of edges in 2D materials [28, 75].

**2.2. The effects of defects**

The defects in 2D materials can have great influence on their properties, such as mechanical, electrical, optical, thermal, and magnetic properties.

The mechanical properties of 2D materials can be strongly affected by the density and the detailed arrangements of defects. A systematic study show that the in-plane Young's modulus of graphene increases with increasing defect density under low vacancy content [76]. For a higher density of vacancies, the elastic modulus decreases. The elastic modulus of graphene is relatively insensitive to sp$^3$-type defects, even at a high density [77]. In addition, the elastic modulus of multilayer graphene is insensitive to high-energy particle irradiation compared to monolayer graphene because of trapping scattered carbon atoms between layers and tending to form interlayer linking around defects, which can partially restore the degraded modulus [78]. Previous work has shown that the strength of GBs can either increase or decrease with the tilt, and the behavior can be explained by continuum mechanics [23]. Figure 2A shows the strength of GBs as a function of tilt angle. GBs in graphene are usually formed by pentagon-heptagon rings. If pentagon-heptagon defects are evenly spaced, the strength of tilt GBs increase as the square of their tilt angles. However, if pentagon-heptagon defects are not evenly spaced, the trend breaks down.

Electronic band structures of 2D materials can be influenced by defects. In TMDs, the states near the band edges primarily have contributions from the metal $d$ states and chalcogen $p$ states. When chalcogen vacancies are created, gap states can be formed in defective TMDs samples [79]. First-principle calculations reveal that, TMDs can be divided into two types: (1) defect-tolerant, semiconductors with a lower tendency to form defect induced deep gap states; (2) the defect-sensitive case with shallow defect levels. In the defect tolerant case, the nature of the bands near the band edges is significantly different, whereas in the defect-sensitive case, they are of mixed nature as shown in Figure 2B. The TMDs based on group VI and X metals form deep gap states upon creation of a chalcogen (S, Se, Te) vacancy, while the TMDs based on group IV metals form only shallow defect levels and are thus predicted to be defect-tolerant.

Electrical properties of 2D materials are highly dependent on defects [80,81]. The presence of monovalent adsorbates or vacancies can set up a limit on mobility of graphene due to scattering, which can be detected by Raman spectroscopy [14]. The GBs have great influence on electrical properties of 2D materials [81]. Resistance of L-R (across domains) is considerably greater at all gate values than L (left) and R (right) for graphene islands with a GB, which shows increased scattering at the GB. As for TMDs, sulphur vacancies exist in $MoS_2$, which introduce localized donor states inside the bandgap and result in hopping transport at room temperature [21]. However, the vacancies can be partially repaired by chemical functionalization of sulphur containing groups and the mobility can be improved shown in Figure 2C [80]. Field effect transistor devices of tilt GBs in $MoS_2$ samples show a decrease in conductance. While devices of the mirror twin samples with channel along the GB show a better conductance than grain devices, and the channel across the GB is same to grain devices [69]. In addition, studies find that 60° GBs with mirror-symmetric are metallic in $MoS_2$ [28], which could provide new functionalities and form intrinsic electronic heterostructures in monolayer $MoS_2$.

The optical properties of TMDs can also be modulated by defects [82]. Previous work has shown that the weak PL of monolayer $MoS_2$ is mostly due to the formation of negative charged excitons (trion) in naturally n-doped sample [83]. However, strong enhancement of PL can be realized by defect engineering of monolayer $MoS_2$, through thermal treatment or oxygen plasma irradiation [33]. The oxygen adsorbed on defect site has very strong bonding energy, which would introduce heavy p doping in $MoS_2$ and hence a conversion from trion to exciton. Defects in TMDs can efficiently localize excitons, which act as sources of single photon emission when suitably isolated [84]. The energy of single photon emission is 20-100meV lower than that of 2D excitons [85] and the emission shows sharp linewidth of 58-500 μeV[86]. Figure 2D-i and ii show PL spectrum of localized emitters and PL mapping of narrow single photon emission lines, respectively. In addition, single photon emission shows two non-degenerate transitions, which are cross-linearly polarized [86].

The thermal transport in graphene can also be modulated by defect engineering, as shown in Figure 2E. Oxygen plasma treatment could reduce the thermal conductivity of graphene even at extremely low defect concentration (83% reduction for ~0.1% defects), which could be attribute mainly to the creation of oxygen containing defects, e.g. carbonyl pair. Other types of defects, such as, hydroxyl, epoxy groups and nano-holes demonstrate much weaker effects on the reduction of thermal conductivity where the $sp^2$ nature of graphene is better preserved [24].

Magnetic properties of 2D materials are also affected by defects. Figure 2F shows that monolayer black phosphorus can have ferromagnetic properties when substitutionally doped with transition metals of V, Cr, Mn, Fe, or Ni [87]. The pristine black phosphorus is nonmagnetic and the magnetization of doped black phosphorus is mainly from the 3d orbitals of the transition metals. In addition, research shows that bare Mo-terminated edge in $MoS_2$ has a ferromagnetic ground state, where every Mo atom at edge possesses a local magnetic moment. However, the Mo magnetic moments are totally quenched after edge reconstruction, while the metallic behavior of the edge is still well-preserved [28].

Besides, electroluminescence (EL) can also be realized at GBs of TMDs under biasing between the two grains as shown in Figure 2G. According to the study of localized light-emission in single carbon nanotubes, defects can create pockets of trapped electrons in SiO$_2$ and forms locally p-doped segments that cause emission under biasing [88,89]. In the case of monolayer WS$_2$, the presence of GBs defects might also lead to locally reside hole donor dopants and result in radiative recombination of the majority electron charge carrier under biasing, namely EL [90].

## 3. Raman spectroscopic investigation of defects

Raman spectroscopy as a nondestructive characterization method is widely used in 2D materials research. It can be adopted to determine the number of layers, monitor the electronic band structure, study the effects of perturbations on 2D materials, such as electric and magnetic fields, strain, doping and functional groups [43, 46, 56, 91]. It has also been frequently used to estimate the nature and numbers of defects in graphene and other 2D materials.

### 3.1 Defect characterization in graphene by Raman spectroscopy

Raman spectroscopy is widely used in graphene researches, as it provides information about both atomic structure and electronic properties of graphene. The Raman spectrum of pristine and defective single layer graphene (SLG) is composed of distinct peaks, including G, D, D′, 2D (also known as G′), 2D′, D + D″ and D + D′ peaks as shown in Figure 3A [45].

The G and 2D peaks located at ~1580 and ~2700 cm$^{-1}$ respectively, are main Raman characteristic lines in Raman spectrum of graphene. The G peak corresponds to the allowed high-frequency $E_{2g}$ phonon (with momentum q≈0) of the Brillouin zone (BZ) at Γ point, and the double-resonance 2D peak is from the $A'_1$ phonon in the BZ corner of K point [46]. A defect is required for the activation of D, D′, and D + D′ peaks. The defect-induced D peak at ~1350 cm$^{-1}$ originates from $A'_1$ phonons(with q≠0) around the Brillouin zone corner K and is activated by intervalley double resonant process, as shown in Figure 3B. The D peak is strongly dispersive with excitation energy [45, 92] as shown in Figure 3C. When the excitation energy increases, the phonons involved in the double resonant process move away from K, and their frequencies become higher [93]. In addition, the Raman intensity of D peak also shows dependency on the excitation energy, $I_D/I_G \sim E_L^{-4}$, as shown in Figure 3D [94,95]. The defect-induced D′ peak at ~1620 cm$^{-1}$ coming from $E_{2g}$ phonon (with $q \neq 0$) is given by intravalley double resonant process that connects two points belonging to the same cone around K (or K′), as shown in Figure 3B. With defect, one intravalley and one intervalley phonon can be emitted together, producing the D + D′ peak at ~2950 cm$^{-1}$. Hence, the intensities of the above mentioned defect-induced Raman peaks (i.e. D, D′, and D + D′) can provide much information on the nature and numbers of defects in graphene.

### 3.1.1 The nature of defects in graphene probed by Raman spectroscopy

Previous studies have shown that, although it is small and usually unnoticed under the noise level, the D peak is generally present in high quality mechanically exfoliated graphene, and typically reaches ~1.5% in amplitude with respect to the G peak ($I_D/I_G$) [14]. Such a weak D peak is present on pristine graphene on different substrates, i.e. $SiO_2$/Si, PMMA, Mica, Glass, and Gold [96]. After graphene is treated by thermal annealing or plasma irradiation, different types of defects, such as vacancies and sp$^3$ hybridization, are introduced [42, 97]. For example, the sp$^3$-defects can be introduced by mild oxidation, and vacancy-defects can be produced by Ar$^+$-bombardment. Figure 4A shows the evolution of Raman spectra of single layer graphene with $H_2$ and Ar$^+$ plasma treatment. With the increase of irradiation time, the D peak increase, together with other defect related Raman peaks, especially, D′ peak. It can be found that graphene samples contain different types of defects have different $I_D/I_{D'}$ ratios.

Theoretical results show that the $I_D/I_{D'}$ ratio can be used for identifying the nature of defects in graphene [98]. The $I_D/I_{D'}$ ratio is determined by the defect potential $\Delta V$, where $\Delta V = V_{DSLG} - V_{SLG}$, $V_{DSLG}$ describes the self-consistent potential of the defective SLG (DSLG) and $V_{SLG}$ is the corresponding value for pristine SLG. $\Delta V$ is different for mono-vacancy (MV), double-vacancy (DV), Stone–Wales (SW), 555-777, and 5555-6-7777 defects in SLG [98]. The $I_D/I_{D'}$ ratios were calculated as ~1 for MV, ~11 for DV, ~17 for SW, ~4 for 555-777, and ~4 for 5555-6-7777 defects, while upon oxygen adsorption the respective values were ~1, ~1, ~10, ~3 and ~6, respectively [98]. To some extent, oxygen adsorption on a MV heals the defect, suppressing the intensity of the D and D′ peaks about two orders of magnitude.

Figure 4B shows that variation of the intensities of D and D′ peaks of different types of defects in experiment. All the $sp^3$-type defective graphene (partially hydrogenated, fluorinated and oxidized graphene) share the same slope in the plot $I_D/I_G$ versus $I_{D'}/I_G$, that is, they have the same $I_D/I_{D'}$ ($\simeq 13$) [100]. In contrast, defective graphene samples produced by ion-bombardment and anodic bonding [101], which mainly contain vacancy-like defects, show a smaller $I_D/I_{D'}$ ($\simeq 7$). While polycrystalline graphite, where defects are commonly GBs, shows an even smaller $I_D/I_{D'}$ ($\simeq 3.5$) [100]. The above results demonstrate that Raman spectroscopy, i.e. $I_D/I_{D'}$, could be used to identify the nature of defects in graphene.

It should be noted that the experimental results are not consistent with theoretical calculations. The discrepancy between theory and experiment can be attributed to the idealized description of defects in the ab initio calculations. Though vacancies defect can be characterized in experiment, there is no difference for MV and DV. During the process of mild oxidation, $sp^3$-defect cannot be described as the only type of defects. The defects are expected to have both on-site and hopping components since the out-of-plane bonding with the atom also introduces distortions in the crystal lattice [99, 102]. Furthermore, defects are usually not isolated (as assumed in the ab initio calculations), but they appear in form of dimers or clusters.

### 3.1.2 Quantifying the numbers of defects in graphene by Raman spectroscopy

Raman spectroscopy can also be used to quantify the amount of defects in graphene [95, 103]. With increasing number of defects produced by $Ar^+$-bombardment, the relationship between the D to G intensity ratio($I_D/I_G$) and average defect distance ($L_D$) undergo two stages as shown in Figure 5A[95, 45]. In stage 1, $I_D$ is almost proportional to the total number of defects probed by the laser spot. For $L_D$ and laser spot size $L_L$, there are on average $(L_L/L_D)^2$ defects in the area probed by the laser, thus $I_D \propto (L_L/L_D)^2$ [95]. On the other hand, $I_G$ is proportional to the total area probed by the laser ($I_G \propto L_L^2$), giving $I_D/I_G \propto 1/L_D^2$ [95]. The D to G intensity ratio reaches a maximum for $L_D \sim 3$ nm. In stage 2, with increasing number of defects, where $L_D < 3$ nm, $I_D$ will decrease with respect to $I_G$ and $I_D/I_G \propto M$ (M being the number of ordered hexagons), and the development of D peak actually indicates the degree of ordering of the system, opposite to stage 1 [95]. This leads to a new relation: $I_D/I_G \propto L_D^2$ [95]. As also mentioned in Figure 3D, $I_D/I_G$, for a specific $L_D$, depends on the laser energy. Figure 5B plots $E_L^4(I_D/I_G)$ as a function of $L_D$, where $E_L$ is excitation energies. For the low-defect density regime (the black line, within stage 1) ($L_D > 10$ nm), after fitting, the defect density is $n_D^2(cm^{-2}) = 7.3 \times 10^{-9} E_L^4(eV^4) \frac{I(D)}{I(G)}$, according to $n_D^2 = 1/(\pi L_D^2)$ [45]. For the high defect density regime (the red line, within stage 2), the defect density is $n_D^2(cm^{-2}) = \frac{5.9 \times 10^{14}}{E_L^4(eV^4)} \left[\frac{I(D)}{I(G)}\right]^{-1}$ [45]. Therefore, the amount of defects in graphene can be effectively estimated by Raman spectroscopy. As defects are scattering or trapping centers for carriers, they could strongly influence the electrical performance of graphene. Figure 5C-i and 5C-ii show Raman spectra and transfer curves of graphene with increasing number of defects introduced by hydrogen plasma, respectively [14]. The results reveal that the $1/\mu$ of graphene has a linear relationship with the ratio $I_D/I_G$ as shown in Figure 5D, which illustrates that intervalley scatters can be as a factor limiting carrier mobility and the scattering probability is proportional to defect density in graphene [14].

### 3.1.3 Identifying edges and grain boundaries in graphene by Raman spectroscopy

Raman spectra can also be used to distinguish the type of edges of graphene, i.e. Zigzag (Z) or Armchair (A) edges [104]. The Raman D peak is inactive for Z-edges because the exchanged momentum by scattering from the Z-edges (dZ) cannot connect the adjacent Dirac cones K and K′ as shown in Figure 6A and hence does not fulfill the double resonant process [105-107]. On the other hand, the exchanged momentum from A-edges (dA) can satisfy the intervalley scattering process between K and adjacent K′ as shown in Figure 6B; hence D peak is Raman active for A-edge. Figure 6C shows Raman images of different edges of mechanically exfoliated single layer graphene sheets [70]. The intensity of D peak for A-edge is higher than that of Z-edge. Because of imperfection of Z-edges, the D peak intensity is indeed not zero. It has also been found that graphene edges (both armchair and zigzag) are not stable and undergo modifications even at temperature as low as 200 ℃ [74]. The D peak also shows its sensitivity in identifying the alignment configuration at edges for $n$LG ($n > 1$) [108].

GBs in graphene can also be treated as a special "edge" and can be monitored by Raman spectroscopy. Figure 6D-i shows the Raman image of two CVD grown graphene domains with a GB, where a pronounced $I_D$ is observed at the GB between two coalesced grains [109]. Hence, Raman image of the D peak intensity provides a convenient way to clearly identify the locations of GBs. Electrical measurement shows that the resistance across the GB is much larger than that within the grain as shown in Figure 6D-ii, reflecting the effect of GBs to impede the electrical transport.

**3.2 Defect characterization in TMDs by Raman spectroscopy**

Raman spectroscopy can also be used to gain insight in the vibrational properties of TMDs, e.g. identifying the number of layers and studying the perturbation induced by strain on their crystal lattice [40, 41]. Monolayer $MoS_2$ has two prominent Raman-active peaks, E′ and A′$_1$. At Γ point, the out-of-plane optical (ZO) branch gives rise to the A′$_1$ peak. The degeneracy of the longitudinal optical (LO) and transverse optical (TO) branches is broken at the zone center due to the slight polarity of $MoS_2$. However, only one peak (E′) is detectable in Raman spectroscopy due to the small LO-TO splitting (<3 cm$^{-1}$) [40,110]. When defects are introduced into $MoS_2$, three changes of Raman spectrum are summarized as follows: (1) the positions of two prominent Raman-active peaks shift, (2) two prominent Raman-active peaks broaden, (3) defect-activated peaks appear.

After introducing vacancy defects into $MoS_2$ by electron-beam irradiation, the redshift of E′ peak and blueshift of the A′$_1$ peak, accompanied by the broadening of E′ and A′$_1$ peaks are observed, as shown in Figure 7A [111, 47]. With the increase of vacancy concentration, there are fewer Mo-S bonds involved in the in-plane vibrations, and thus the restoring force constant of the E′ peak is continuously weakened, resulting in the redshift of E′ peak. The A′$_1$ peak corresponds to out-of-plane vibrations of the Mo-S bonds with static center Mo atoms for the pristine system. Its restoring force constant is slightly reduced because of missing a Mo-S bond. However, the originally static Mo atom is allowed to vibrate out of plane, which strengthens the restoring force constant from the Mo-S bond vibration, resulting in the blueshift of A′$_1$ [111]. Figure 7B presents the frequency differences between E′ and A$_1$′ peaks with the increase of defect concentration, which shows an almost linear dependence. During the process of low energy phosphorus implantation, significant broadening of E′ peak in the MoS$_2$ flake was also observed [27].

Figure 7C shows the evolution of Raman spectra of monolayer MoS$_2$ as a function of L$_D$, where L$_D$ is an average interdefect distance, which can be used for expressing the defect density. The L$_D$ of vacancy defects in monolayer MoS$_2$ flakes can be tuned by Mn+ bombardment [47] with the relationship L$_D$ = 1/$\sqrt{\sigma}$, where $\sigma$ is the ion dose density. As can be seen, several defect-activated peaks appear in MoS$_2$ samples. These peaks involve phonons at the zone edge of the BZ, which may be activated by the momentum contribution from the defect, allowing the Raman selection rule to be satisfied. The most intense peak is located at ∼227 cm$^{-1}$, which is attributed to disorder-induced Raman scattering and has been assigned to LA phonons with momentum q≠0 at M point. The evolution of the intensities (peak height) of the LA peak normalized to the E′ and A′$_1$ peaks, as a function of L$_D$ is shown in Figure 7D. Fitting the data in Figure 7D using $\frac{I(LA)}{I(X)} = \frac{C(X)}{L_D^2}$, where X = E′ or A′$_1$, reveals that C (E′) = 1.11 ±0.08 nm$^2$ and C (A′$_1$) = 0.59 ±0.03 nm$^2$ [47].

In addition to vacancy defects, oxygen substitution in TMDs can also be characterized by Raman spectroscopy. The O substitution can partially restore the bonding but cannot fully restore the Raman peak positions of the pristine system, and would be treated as a vacancy from the Raman peak shifts. In WSe$_2$ sample, two features located at ~260 and ~263 cm$^{-1}$ might correspond to the overtone of the LA phonon branch at the M point and to a phonon belonging to the A-symmetry optic branch at the M point, which becomes active due to structural disorder [112]. After air heating at 400 ℃, two peaks at 695 and 803 cm$^{-1}$ are observed for few-layer WSe$_2$ sample, which are assigned to tungsten oxide (WO$_{3-x}$) [113].

The above results reveal that Raman spectroscopy can be used to study the defects in TMDs. However, it is found that the change of Raman features with the increase of defect density is rather insensitive. This is different from the case of graphene, where the D peak originates from the double resonant Raman process and is very sensitive the defect density. Following, we will show that PL spectroscopy provide a more reliable and sensitive way to monitor the defects in TMDs.

## 4. Defect characterization in TMDs by PL spectroscopy

In TMDs, the presence of vacancy will introduce defect density of states within the bandgap [21]. Figure 8A shows band structure and density of states of monolayer MoS$_2$ with S vacancies. The bottom of the conduction band of MoS$_2$ is dominated by Mo 4d orbitals, while top of valence band is originated from the hybridization between strong Mo 4d orbitals and weak S 3p orbitals. Therefore, midgap states can be formed after introducing S vacancies in MoS$_2$ and strong localization can be formed surrounding the vacancy [21]. At room temperature, monolayer TMDs samples have strong PL signals due to the direct bandgap emissions: the coulomb interaction between one electron and one hole creates exciton (X$_0$), and exciton further charged by binding an additional electron or hole to form charged three-body excitons named as trions (X$_0^-$ or X$_0^+$) [114]. At low temperature, in addition to exciton and trion emissions, there is another PL peak attributed to emission from bound exciton, named X$_b$, which results from exciton bound to midgap states (localized states) within the bandgap [115]. The localized states formed within the bandgap of TMDs can be attributed to lattice defects (such as the vacancy as shown in Figure 8A) or residual impurities introduced during the mechanical exfoliation process or ion irradiation [115-118].

Four main features of bound exciton $X_b$ peak have been concluded as following: 1. a nonlinear laser power dependence and saturation phenomena at high laser power; 2. vanished with increasing temperature; 3. circular polarization dependence; 4. electrical gate dependence. Figure 8B shows PL spectra of $MoS_2$ measured at 80K and the intensity of $X_0$ peak is normalized by $X_b$. It is found that the relationship between PL intensity (I) and excitation power (p) follows the power law $I \propto p^k$ with k<1 as shown in Figure 8C. The nonlinear laser power dependence and saturation phenomena for $X_b$ peak can be explained by the population of defect states with excitons at high excitation power [119]. The increasing excitation intensity will also cause band filling of the localized energy states, giving rise to the blueshifts of $X_b$ peak. The PL intensity of $X_b$ peak is vanished with increasing temperature as shown in Figure 8D, which is understandable since excitons are not tightly bound to defects and such weak interaction can be easily perturbed by thermal stimulation [120]. PL emissions of neutral and charged excitonic states (exciton and trion emissions) are highly circularly polarized due to valley optical selection rules derived from the single particle picture [116]. The bound exciton emission also shows a small circular polarization of 13% at a temperature of $T$ = 10 K for $WSe_2$ sample as shown in Figure 8E [116]. The degree of circular polarization is given by: $\rho = \frac{I(\sigma^+)-I(\sigma^-)}{I(\sigma^+)+I(\sigma^-)}$, where $I(\sigma^+)$ and $I(\sigma^-)$ correspond to the PL intensity of the $\sigma^+$ and $\sigma^-$ polarization components, respectively [116]. One possible mechanism could be related to partial transfer of the valley polarization from optically generated electron-hole pairs [121] to the localized electrons or holes [116]. The electrical gate dependence of PL spectra of $MoS_2$ at 10K is shown in Figure 8F. With gate voltage Vg varying from 60 V to -60 V, the intensity of $X_0$ peak at ~1.958 eV increases while the intensity of $X_0^-$ peak at ~1.925 eV decreases, which is accompanied by the change from electron doping to hole doping in $MoS_2$. Since $X_b$ peak results from exciton bound to localized states, its intensity also shows strong gate dependence and increases with Vg varying from 60 V to -60 V, similar to $X_0$ peak.

**4.1 Quantifying the numbers of defects in TMDs by PL spectroscopy**

Following, we will show that the intensity of bound exciton $X_b$ peak can be used to quantify the number of defects in TMDs. Figure 9A shows low temperature PL spectra (80 K) of single layer $WSe_2$ with and without electron-beam irradiation during the electron beam lithography (EBL) process [48]. As can be seen, a strong $X_b$ peak presents in the PL spectrum of electron-beam irradiated region, which suggests that PL spectroscopy can be used to monitor the structural defects in TMDs introduced by electron-beam irradiation. Figure 9B shows temperature dependence of PL spectra of electron-beam irradiated $WSe_2$. A thermally dissociation process can be used to describe the population of $X_b$, that is $N_{Xb}$, as a function of temperature T:

$$N_{xb}(T) = \frac{N_0}{1+(\tau/\tau_0)e^{-E_A/kT}}$$, where $\tau$ is the excitonic lifetime exceeding 100ps, $\tau_0$ is effective scattering time and $E_A$ is activation energy [48]. By fitting the data in Figure 9C with above equation, we get $E_A$ of 43 meV and the ratio $\tau/\tau_0$ of 259 for bound exciton in monolayer $WSe_2$. Figure 9D shows the PL spectra of electron-beam irradiated $WSe_2$ with different electron dosage. The $X_b$ intensities are plotted against laser power with different electron irradiation density as shown in Figure 9E. An obvious sublinear dependence can be observed and can be well fitted by power law $I \propto P^k$, where k is ~0.59. The saturated intensity of $X_b$ increases with increasing electron-beam density, in good accordance with the increase of defect densities at higher dosage. In Figure 9F, in order to quantify defects, the intensity ratio of $I_{Xb}/I_{X0}$ is obtained, where $X_0$ peak was used for normalization, similar to defect characterization in graphene by Raman intensity ratio of D and G peaks ($I_D/I_G$) [14]. It can be seen that such ratio shows very good linear dependence with the irradiation electron dosage in the range of $<60 \times 10^6 \mu m^{-2}$, which suggest that the intensity of $X_b$ peak can be used as a standard approach to characterize and monitor the defects in $WSe_2$ sample [48]. The electrical properties of $WSe_2$ can be greatly influenced by the

presence of defects, as shown in Figure 9G. It is interesting to find that $1/\mu$ increases almost linearly with the increase of irradiation electron dosage and also $I_{Xb}/I_{X0}$. The defects, e.g. vacancies, are efficient short-range scatters for conducting carriers, which would result in $1/\mu \propto N_{sr}$, where $N_{sr}$ is the number of scattering centers or defects and presented by the electron irradiation dosage or $I_{Xb}/I_{X0}$. The above results clear demonstrate that PL spectroscopy is a nondestructive and efficient method to investigate the defects in TMDs.

**4.2 Edge and grain boundary in TMDs probed by PL spectroscopy**

With the development of CVD techniques, large-area monolayer TMDs crystals are available, while spatial nonuniformities in the PL intensity from TMDs grains are frequently observed at room temperature [69, 122, 123]. GB and edges always exhibit PL signals different from that at the crystal center. These nonuniformities can be attributed to localized states, doping, and strain [69, 78, 123, 124].

Figure 10A and 10B shows PL and Raman images of monolayer $MoS_2$ after thermal annealing. Strong PL enhancement is observed in some regions which has lower Raman intensity. Figure 10C shows a clear blueshift of the $MoS_2$ $A'_1$ peak in these regions. AFM measurement (Figure 10D) reals that PL enhancement is from cracks that formed during annealing, where oxygen molecule chemically bound to defect site. The charge transfer from $MoS_2$ to oxygen would induce a trion to exciton transformation and hence the PL enhancement [33]. Furthermore, excitons may localize at the defect/crack sites, which would have much larger binding energy and suppress the thermally activated non-radiative recombination, result in a very high PL quantum efficiency. There are also results demonstrate that higher PL intensity and lower-energy PL emission present at the edges and GBs of $WS_2$, which is consistent with stronger exciton binding at localized states [83]. The examined edges of $WS_2$, such as bare sulfur, bare tungsten, sulfur, or oxygen passivated tungsten, support edge-localized states at the Fermi level [83]. It is noteworthy that sulfur passivation has been studied in zigzag $MoS_2$ triangular islands using STM, identifying states at

Fermi level at the edges [125]. Scanning transmission electron microscope (STEM) studies on GBs in 2D TMDs have found that a variety of non-6-membered rings (4-, 5-, 7- and 8-membered rings) together with strained 6-membered rings are present at GBs with different misorientation angles [124]. DFT calculation has shown that all these non-6-membered dislocation cores can induce deep gap states [124].

Studies also found that tilt and mirror GBs have different PL intensity in TMDs as shown in Figure 10 E and F. The mirror boundary is with quenching of PL intensity and a 8 meV blueshift in peak energy, while the tilt boundary shows a surprising enhancement in emission strength and a 26 meV blueshift. The PL spectrum of the GBs may be affected by two main factors: doping and strain. The defects in the mirror boundary are molybdenum rich, which would n-dope the boundary; whereas the defects in the tilt boundary are sulphur rich, which would p-dope the boundary [69]. Since PL intensity is strongly affected by charge density [83], the increased/decreased electron density at mirror/tilt edges would cause PL quenching/enhancement [69]. On the other hand, strain around the boundaries may modify the bandgap or cause the boundary region to lift off the electrically disordered $SiO_2$ surface, hence enhances PL emission [69].

## 5. Conclusions and perspective

In this review, we have classified the types of defects in 2D materials, represented by graphene and TMDs, based on dimensionality and atomic structure. These structural defects have been thoroughly studied by modern characterization techniques, such as TEM, STM, and XPS. Optical spectroscopic methods, for example Raman and PL, are time efficient and nondestructive techniques for defect characterization in 2D materials. Defects in graphene can be well classified and quantified by Raman spectroscopy. The intensity ratio of D and D′ peaks can be used for identifying the types of defects and the intensity ratio of D and G peaks can be used for quantifying the numbers of defects. In the case of TMDs, we have shown that PL spectroscopy can be used to characterize the defects inside. The intensity of low temperature bound exciton emission $X_b$ is very sensitive to the number of defects in

TMDs. GB can also be identified by PL spectroscopy and enhanced PL is presented at edges because of localized states, doping and strain. Although great progress on defect characterization by optical spectroscopy has been made, more efforts are expected on but not restricted to the following aspects:

1. Though quantification of defects in TMDs has been done by PL spectroscopy, the characterization of the types of defects by optical spectroscopic methods has not been reported. Different types of defects can form different localized states in 2D materials, which would give bound exciton emission at different energies.

2. Modulation of PL intensity has been achieved by defect engineering in 2D materials. Recently, single photon emitter has been observed in mechanically exfoliated and CVD grown TMDs. It is possible to utilize defect engineering to further modulate single photon emission in TMDs.

3. The defects have great influence on properties of 2D materials, such as mechanical, electrical, optical, thermal, and magnetic properties. More work should be carried out on the application of defects in tuning the properties of TMDs devices, such as, EL, photodetector, and so on. For example, although EL at GBs has been observed, there are other types of defects which could contribute to the improvement of EL quantum efficiency. Vacancies or adatoms can be introduced by plasma treatment, which may form center for radiative recombination of EL.

4. This paper mainly focuses on studies of defects in graphene and TMDs. Spectroscopic investigation should also be applied for defects in other 2D materials, such as black phosphorus, $Bi_2Se_3$, silicene and so on.

**Acknowledgements**

This work is supported by NSFC (61422503 and 61376104), the open research funds of Key Laboratory of MEMS of Ministry of Education (SEU, China), and the Fundamental Research Funds for the Central Universities.

**References**

[1] Geim AK. Graphene: Status and Prospects. Science 2009, 324, 1530-1534.

[2] Balandin AA, Ghosh S, Bao WZ, et al. Superior Thermal Conductivity of Single-Layer Graphene. Nano Lett 2008, 8, 902-907.

[3] Du X, Skachko I, Barker A, Andrei EY. Approaching ballistic transport in suspended graphene. Nat Nanotechnol 2008, 3, 491-495.

[4] Radisavljevic B, Radenovic A, Brivio J, Giacometti V, Kis A. Single-layer $MoS_2$ transistors. Nat Nanotechnol 2011, 6,147–150.

[5] Dean CR, Young AF, Meric I, et al. Boron nitride substrates for high-quality graphene electronics. Nat Nanotechnol 2010, 5, 722–726.

[6] Li LK, Yu YJ, Ye GJ, et al. Black phosphorus field-effect transistors. Nat Nanotechnol 2014, 9, 372–377.

[7] Mak KF, Lee C, Hone J, Shan J, Heinz TF. Atomically thin $MoS_2$: a new direct-gap semiconductor. Phys Rev Lett 2010, 105, 136805.

[8] Wang QH, Kalantar-Zadeh K, Kis A, Coleman JN, Strano MS. Electronics and optoelectronics of two-dimensional transition metal dichalcogenides. Nat Nanotechnol 2012, 7, 699-712.

[9] Eda G, Maier SA. Two-dimensional crystals: managing light for optoelectronics. ACS Nano 2013, 7, 5660-5665.

[10] Lembke D, Kis A. Breakdown of high-performance monolayer $MoS_2$ transistors. ACS Nano 2012, 6, 10070-10075.

[11] Zhou HL, Wang C, Shaw JC, et al. Large area growth and electrical properties of p-type $WSe_2$ atomic layers. Nano Lett 2015, 15, 709-713.

[12] Lopez-Sanchez O, Lembke D, Kayci M, Radenovic A, Kis A. Ultrasensitive photodetectors based on monolayer $MoS_2$. Nat Nanotechnol 2013, 8, 497-501.

[13] Banhart F, Kotakoski J, Krasheninnikov AV. Structural Defects in Graphene. ACS Nano 2011, 5, 26-41.

[14] Ni ZH, Ponomarenko LA, Nair RR, et al. On Resonant Scatterers As a Factor Limiting Carrier Mobility in Graphene. Nano Lett 2010, 10, 3868.

[15] Li XS, Cai WW, An J, et al. Large-Area Synthesis of High-Quality and Uniform

Graphene Films on Copper Foils. Science 2009, 324, 1312-1314.

[16] Novoselov KS, Geim AK, Morozov SV, et al. Electric Field Effect in Atomically Thin Carbon Films. Science 2004, 306, 666-669.

[17] Novoselov KS, Geim AK, Morozov SV, et al. Two-dimensional gas of massless Dirac fermions in graphene. Nature 2005, 438, 197-200.

[18] Novoselov KS, Jiang D, Schedin F, et al. Two-Dimensional Atomic Crystals. Proc Natl Acad Sci USA 2005, 102, 10451-10453.

[19] Zhang YB, Tan YW, Stormer HL, Kim P. Experimental observation of the quantum Hall effect and Berry's phase in graphene. Nature 2005, 438, 201-204.

[20] Hong JH, Hu ZX, Probert M, et al. Exploring atomic defects in molybdenum disulphide monolayers. Nat Commun 2015, 6, 6293.

[21] Qiu H, Xu T, Wang ZL, et al. Hopping transport through defect-induced localized states in molybdenum disulphide. Nat Commun 2013, 4, 2642.

[22] Lee GH, Cooper RC, An SJ, et al. High-Strength Chemical-Vapor–Deposited Graphene and Grain Boundaries. Science 2013, 340, 1073-1076.

[23] Wei YJ, Wu JT, Yin HQ, Shi XH, Yang RG, Dresselhaus M. The nature of strength enhancement and weakening by pentagon–heptagon defects in graphene. Nat Mater 2012, 11, 759-763.

[24] Zhao WW, Wang YL, Wu ZT, et al. Defect-Engineered Heat Transport in Graphene: A Route to High Efficient Thermal Rectification. Sci Rep 2015, 5, 11962.

[25] Carr LD, Lusk MT. Defect Engineering: Graphene Gets Designer Defects. Nat Nanotechnol 2010, 5, 316-317.

[26] Boukhvalov DW, Katsnelson MI. Chemical functionalization of graphene with defects. Nano Lett 2008, 8, 4373-4379.

[27] Nipane A, Karmakar D, Kaushik N, Karande S, Lodha S. Few-Layer $MoS_2$ p‑Type Devices Enabled by Selective Doping Using Low Energy Phosphorus Implantation. ACS Nano 2016, 10, 2128-2137.

[28] Zhou W, Zou XL, Najmaei S, et al. Intrinsic Structural Defects in Monolayer Molybdenum Disulfide. Nano Lett 2013, 13, 2615-2622.

[29] Li Z, Yang S, Dhall R, et al. Layer Control of $WSe_2$ via Selective Surface Layer Oxidation. ACS Nano 2016, 10, 6836-6842.

[30] Kim E, Ko C. Kim K, Chen Y, et al. Site Selective Doping of Ultrathin Metal Dichalcogenides by Laser-Assisted Reaction. Adv Mater 2016, 28, 341-346.

[31] Rong YM, Sheng YW, Pacios M, et al. Electroluminescence Dynamics across Grain Boundary Regions of Monolayer Tungsten Disulfide. ACS Nano 2016, 10, 1093-1100.

[32] Kibsgaard J, Chen Z, Reinecke BN, Jaramillo TF. Engineering the surface structure of $MoS_2$ to preferentially expose active edge sites for electrocatalysis. Nat Mater 2012, 11, 963-969.

[33] Nan HY, Wang ZL, Wang WH, et al. Strong photoluminescence enhancement of $MoS_2$ through defect engineering and oxygen bonding. ACS Nano 2014, 8, 5738-5745.

[34] Lin YC, Bjorkman T, Komsa HP, et al. Three-fold rotational defects in two-dimensional transition metal dichalcogenides. Nat Commun 2015, 6, 6736.

[35] Lin, JH, Pantelides ST, Zhou W. Vacancy-induced formation and growth of inversion domains in transition-metal dichalcogenide monolayer. ACS Nano 2015, 9, 5189-5197.

[36] Addou R, Mcdonnell S, Barrera D, et al. Impurities and electronic property variations of natural $MoS_2$ crystal surfaces. ACS Nano 2015, 9, 9124-9133.

[37] Fuhr JD, Saul A, Sofo JO. Scanning tunneling microscopy chemical signature of point defects on the $MoS_2(0001)$ surface. Phys Rev lett 2004, 92, 026802.

[38] Li H, Tsai C, Koh AL, et al. Activating and optimizing $MoS_2$ basal planes for hydrogen evolution through the formation of strained sulphur vacancies. Nat Mater 2016, 15, 48.

[39] Zhang X, Tan QH, Wu JB, Shi W, Tan PH. Review on the Raman spectroscopy of different types of layered materials. Nanoscale 2016, 8, 6435-6450.

[40] Conley HJ, Wang B, Ziegler JI, HaglundJr RF, Pantelides ST, Bolotin KI. Bandgap engineering of strained monolayer and bilayer $MoS_2$. Nano Lett 2013, 13, 3626-3630.

[41] Liu YL, Nan HY, Wu X, et al. Layer-by-layer thinning of $MoS_2$ by plasma. ACS Nano 2013, 7, 4202-4209.

[42] Nan HY, Ni ZH, Wang J, Zafar Z, Shi ZX, Wang YY. The thermal stability of graphene in air investigated by Raman spectroscopy. J Raman Spectrosc 2013, 44, 1018-1021.

[43] Chen YM, Meng LJ, Zhao WW, et al. Raman mapping investigation of chemical vapor deposition-fabricated twisted bilayer graphene with irregular grains. Phys Chem Chem Phy 2014, 16, 21682-21687.

[44] Tuinstra F, Koenig JL. Raman spectrum of graphite. J Chem Phys 1970, 53, 1126-1130.

[45] Ferrari AC, Basko DM. Raman spectroscopy as a versatile tool for studying the properties of graphene. Nat nanotechnol 2013, 8, 235-246.

[46] Ferrari AC, Meyer JC, Scardaci V, et al. Raman spectrum of graphene and graphene layers. Phys Rev Lett 2006; 97, 187401.

[47] Mignuzzi S, Pollard AJ, Bonini N, et al. Effect of disorder on Raman scattering of single-layer $MoS_2$. Phys Rev B 2015, 91, 195411.

[48] Wu ZT, Luo ZZ, Shen YT, et al. Defects as a factor limiting carrier mobility in $WSe_2$: a spectroscopic investigation. arXiv preprint arXiv:1608.02043, 2016.

[49] Stone AJ, Wales DJ. Theoretical studies of icosahedral $C_{60}$ and some related species. Chem Phys Lett 1986, 128, 501-503.

[50] Ma J, Alfe D, Michaelides A, Wang E. Stone-Wales defects in graphene and other planar sp2-bonded materials. Phys Rev B 2009, 80, 033407.

[51] Meyer JC, Kisielowski C, Erni R, Rossell MD, Crommie MF, Zettl A. Direct imaging of lattice atoms and topological defects in graphene membranes. Nano Lett 2008, 8, 3582-3586.

[52] Kotakoski J, Krasheninnikov AV, Kaiser U, Meyer JC. From point defects in graphene to two-dimensional amorphous carbon. Phys Rev Lett 2011, 106, 105505.

[53] Lee GD, Wang CZ, Yoon E, Hwang NM, Kim DY, Ho KM. Diffusion, coalescence, and reconstruction of vacancy defects in graphene layers. Phys Rev

Lett 2005, 95, 205501.

[54] Tongay S, Zhou J, Ataca C, et al. Broad-range modulation of light emission in two-dimensional semiconductors by molecular physisorption gating. Nano Lett 2013, 13, 2831-2836.

[55] Choi J, Zhang H, Choi JH. Modulating Optoelectronic Properties of Two-Dimensional Transition Metal Dichalcogenide Semiconductors by Photoinduced Charge Transfer. ACS Nano 2016, 10, 1671-1680.

[56] Wu ZT, Zhao WW, Chen WY, et al. The influence of chemical solvents on the properties of CVD graphene. J Raman Spectrosc 2015, 46, 21-24.

[57] Cretu O, Krasheninnikov AV, Rodrı́guez-Manzo JA, Sun L, Nieminen R, Banhart F. Migration and Localization of Metal Atoms on Graphene. Phys Rev Lett 2010, 105, 196102.

[58] Eckmann A, Felten A, Mishchenko A, et al. Probing the nature of defects in graphene by Raman spectroscopy. Nano Lett 2012, 12, 3925-3930.

[59] Komsa HP, Krasheninnikov AV. Native defects in bulk and monolayer $MoS_2$ from first principles. Phys Rev B 2015, 91, 125304

[60] Lu J, Carvalho A, Chan XK, et al. Atomic healing of defects in transition metal dichalcogenides. Nano Lett 2015, 15, 3524-3532.

[61] Nemec N, Tománek D, Cuniberti G. Contact dependence of carrier injection in carbon nanotubes: an ab initio study. Phys Rev Lett 2006, 96, 076802.

[62] Ci LJ, Song L, Jin CH, et al. Atomic layers of hybridized boron nitride and graphene domains. Nat Mater 2010, 9, 430-435.

[63] Zafar Z, Ni ZH, Wu X, et al. Evolution of Raman spectra in nitrogen doped graphene. Carbon 2013, 61, 57-62.

[64] Komsa HP, Kotakoski J, Kurasch S, Lehtinen O, Kaiser U, Krasheninnikov AV. Two-dimensional transition metal dichalcogenides under electron irradiation: defect production and doping. Phys Rev Lett 2012, 109, 035503.

[65] Jeong BW, Ihm J, Lee GD. Stability of dislocation defect with two pentagon-heptagon pairs in graphene. Phys Rev B 2008, 78, 165403.

[66] Komsa HP, Kurasch S, Lehtinen O, Kaiser U, Krasheninnikov AV. From point to

extended defects in two-dimensional MoS$_2$: evolution of atomic structure under electron irradiation. Phys Rev B 2013, 88, 035301.

[67] Huang PY, Ruiz-Vargas CS, van der Zande AM, et al. Grains and grain boundaries in single-layer graphene atomic patchwork quilts. Nature 2011, 469, 389-392.

[68] Zou X, Liu Y, Yakobson BI. Predicting dislocations and grain boundaries in two-dimensional metal-disulfides from the first principles. Nano Lett 2012, 13, 253-258.

[69] van der Zande AM, Huang PY, Chenet DA, et al. Grains and grain boundaries in highly crystalline monolayer molybdenum disulphide. Nat Mater 2013, 12, 554-561.

[70] You YM, Ni ZH, Yu T, Shen ZX. Edge chirality determination of graphene by Raman spectroscopy. Appl Phys Lett 2008, 93, 3112.

[71] Cao D, Shen T, Liang P, Chen XS, Shu HB. Role of Chemical Potential in Flake Shape and Edge Properties of Monolayer MoS$_2$. J Phys Chem C 2015, 119, 4294-4301.

[72] Schweiger H, Raybaud P, Kresse G, Toulhoat H. Shape and edge sites modifications of MoS$_2$ catalytic nanoparticles induced by working conditions: A theoretical study. J Catal 2002, 207, 76-87.

[73] Chen JY, Liu B, Liu YP, et al. Chemical Vapor Deposition of Large-Sized Hexagonal WSe$_2$ Crystals on Dielectric Substrates. Adv Mater 2015, 27, 6722-6727.

[74] Xu YN, Zhan D, Liu L, et al. Thermal dynamics of graphene edges investigated by polarized Raman spectroscopy. ACS Nano 2010, 5, 147-152.

[75] Li YF, Zhou Z, Zhang SB, Chen ZF. MoS$_2$ nanoribbons: high stability and unusual electronic and magnetic properties. J Am Chem Soc 2008, 130,16739-16744.

[76] López-Polín G, Gómez-Navarro C, Parente V, et al. Increasing the elastic modulus of graphene by controlled defect creation. Nat Phys 2015, 11, 26-31.

[77] Zandiatashbar A, Lee GH, An SJ, et al. Effect of defects on the intrinsic strength

and stiffness of graphene. Nat commun 2014, 5.

[78] Liu K, Hsin CL, Fu D, et al. Self‐Passivation of Defects: Effects of High‐Energy Particle Irradiation on the Elastic Modulus of Multilayer Graphene. Adv Mater 2015, 27, 6841-6847.

[79] Pandey M, Rasmussen FA, Kuhar K, Olsen T, Jacobsen KW, Thygesen KS. Defect-Tolerant Monolayer Transition Metal Dichalcogenides. Nano Lett 2016, 16, 2234-2239.

[80] Yu Z, Pan Y, Shen Y, et al. Towards intrinsic charge transport in monolayer molybdenum disulfide by defect and interface engineering. Nat commun 2014, 5.

[81] Tsen AW, Brown L, Levendorf MP, et al. Tailoring electrical transport across grain boundaries in polycrystalline graphene. Science 2012, 336, 1143-1146.

[82] Liu H, Lu J, Ho K, et al. Fluorescence Concentric Triangles: A Case of Chemical Heterogeneity in WS2 Atomic Monolayer. Nano Lett 2016, 16, 5559-5567.

[83] Mak KF, He K, Lee C, et al. Tightly bound trions in monolayer $MoS_2$. Nat Mater 2013, 12, 207-211.

[84] Koperski M, Nogajewski K, Arora A, Marcus J, Kossacki P, Potemsk M. Single photon emitters in exfoliated $WSe_2$ structures. Nat Nanotechnol 2015, 10, 503-506.

[85] Srivastava A, Sidler M, Allain AV, Lembke DS, Kis A, Imamoglu A. Optically active quantum dots in monolayer $WSe_2$. Nat Nanotechnol 2015, 10, 491-496.

[86] He YM, Clark G, Schaibley JR, et al. Single quantum emitters in monolayer semiconductors. Nat Nanotechnol 2015, 10, 497-502.

[87] Wang YR, Pham A, Li S, Yi JB. Electronic and Magnetic Properties of Transition-Metal-Doped Monolayer Black Phosphorus by Defect Engineering. J Phys Chem C 2016, 120, 9773-9779.

[88] Freitag M, Tsang JC, Kirtley J, et al. Electrically Excited, Localized Infrared Emission from Single Carbon Nanotubes. Nano Lett 2006, 6, 1425−1433.

[89] Avouris P, Freitag M, Perebeinos V. Carbon-Nanotube Photonics and Optoelectronics. Nat Photonics 2008, 2, 341−350.

[90] Rong Y, Sheng Y, Pacios M, et al. Electroluminescence Dynamics across Grain

Boundary Regions of Monolayer Tungsten Disulfide. ACS Nano 2015, 10, 1093-1100.

[91] Ni ZH, Yu T, Lu YH, Wang YY, Feng YP, Shen ZX. Uniaxial strain on graphene: Raman spectroscopy study and band-gap opening. ACS Nano 2008, 2, 2301-2305.

[92] Thomsen C, Reich S. Double resonant Raman scattering in graphite. Phys Rev Lett 2000, 85, 5214.

[93] Venezuela P, Lazzeri M, Mauri F. Theory of double-resonant Raman spectra in graphene: intensity and line shape of defect-induced and two-phonon bands. PhysRev B 2011, 84, 035433.

[94] Rodriguez-Nieva JF, Barros EB, Saito R, Dresselhaus MS. Disorder-induced double resonant Raman process in graphene. Phys Rev B 2014, 90, 235410.

[95] Cançado LG, Jorio A, Ferreira EH M, et al. Quantifying defects in graphene via Raman spectroscopy at different excitation energies. Nano Lett 2011, 11, 3190-3196.

[96] Ponomarenko LA, Yang R, Mohiuddin TM, et al. Effect of a high-κ environment on charge carrier mobility in graphene. Phys Rev Lett 2009, 102, 206603.

[97] Guo XT, Zafar A, Nan HY, et al. Manipulating fluorescence quenching efficiency of graphene by defect engineering. Appl Phys Express 2016, 9, 055502.

[98] Jiang J, Pachter R, Mehmood F, Islam AE, Maruyama B, Boeckl JJ. A Raman spectroscopy signature for characterizing defective single-layer graphene: Defect-induced I (D)/I (D′) intensity ratio by theoretical analysis. Carbon 2015, 90, 53-62.

[99] Venezuela P, Lazzeri M, Mauri F. Theory of double-resonant Raman spectra in graphene: intensity and line shape of defect-induced and two-phonon bands. Phys Rev B 2011, 84, 035433.

[100] Eckmann A, Felten A, Mishchenko A, et al. Probing the nature of defects in graphene by Raman spectroscopy. Nano Lett 2012, 12, 3925-3930.

[101] Moldt T, Eckmann A, Klar P, et al. High-yield production and transfer of graphene flakes obtained by anodic bonding. ACS Nano 2011, 5, 7700-7706.

[102] Nair R R, Ren W, Jalil R, et al. Fluorographene: A Two-Dimensional Counterpart of Teflon. Small 2010, 6, 2877-2884.

[103] Lucchese MM, Stavale F, Ferreira EH M, et al. Quantifying ion-induced defects and Raman relaxation length in graphene. Carbon 2010, 48, 1592-1597.

[104] Casiraghi C, Hartschuh A, Qian H, et al. Raman spectroscopy of graphene edges. Nano Lett 2009, 9, 1433-1441.

[105] Pimenta MA, Dresselhaus G, Dresselhaus MS, Cancado LG, Jorio A, Saito R. Studying Disorder in Graphite-Based Systems by Raman Spectroscopy. Phys Chem Chem Phys 2007, 9, 1276–1291.

[106] Saito R, Jorio A, Souza AG, Dresselhaus G, Dresselhaus MS, Pimenta MA. Probing Phonon Dispersion Relations of Graphite by Double Resonance Raman Scattering. Phys Rev Lett 2002, 88, 027401.

[107] Thomsen C, Reich S. Doable Resonant Raman Scattering in Graphite. Phys Rev Lett 2000, 85, 5214–5217.

[108] Li QQ, Zhang X, Han WP, et al. Raman spectroscopy at the edges of multilayer graphene. Carbon 2015, 85: 221-224.

[109] Yu Q, Jauregui LA, Wu W, et al. Control and characterization of individual grains and grain boundaries in graphene grown by chemical vapour deposition. Nat mater 2011, 10, 443-449.

[110] Rice C, Young RJ, Zan R, Bangert U. Raman-scattering measurements and first-principles calculations of strain-induced phonon shifts in monolayer $MoS_2$. Phys Rev B 2013, 87, 081307.

[111] Parkin WM, Balan A, Liang L, et al. Raman shifts in electron-irradiated monolayer $MoS_2$. ACS Nano, 2016, 10, 4134-4142.

[112] del Corro E, Terrones H, Elias A, et al. Excited excitonic states in 1L, 2L, 3L, and bulk $WSe_2$ observed by resonant Raman spectroscopy. ACS Nano 2014, 8, 9629-9635.

[113] Windom BC, Sawyer WG, Hahn DW. A Raman spectroscopic study of $MoS_2$ and MoO3: applications to tribological systems. Triboly Lett 2011, 42, 301-310.

[114] Ross JS, Wu S, Yu H, et al. Electrical control of neutral and charged excitons in

a monolayer semiconductor. Nat Commun 2013, 4, 1474.

[115] Tongay S, Suh J, Ataca C, et al. Defects activated photoluminescence in two-dimensional semiconductors: interplay between bound, charged, and free excitons. Sci Rep 2013, 3, 2657.

[116] Huang J, Hoang TB, Mikkelsen MH. Probing the origin of excitonic states in monolayer WSe$_2$. Sci Rep 2016, 6, 22414.

[117] Yan TF, Qiao XF, Liu XN, Tan PH, Zhang XH. Photoluminescence properties and exciton dynamics in monolayer WSe$_2$. Appl Phys Lett 2014, 105, 101901.

[118] Pelant I, Valenta J. Luminescence Spectroscopy of Semiconductors. Oxford Scholarship, 2012, 180–181.

[119] Schmidt T, Lischka K, Zulehner W. Excitation-power dependence of the near-band-edge photoluminescence of semiconductors. Phys Rev B 1992, 45, 8989-8994.

[120] Bacher G, Schweizer H, Kovac J, Forchel A. Influence of barrier height on carrier dynamics in strained InxGa1-xAs/GaAs quantum wells. Phys Rev B 1991, 43, 9312-9315.

[121] Jones AM, Yu H, Ghimire NJ, et al. Optical generation of excitonic valley coherence in monolayer WSe$_2$. Nat nanotechnol 2013, 8, 634-638.

[122] Cong C, Shang J, Wu X, et al. Synthesis and Optical Properties of Large-Area Single-Crystalline 2D Semiconductor WS$_2$ Monolayer from Chemical Vapor Deposition. Adv Opt Mater 2014, 2, 131-136.

[123] Gutiérrez HR, Perea-López N, Elías AL, et al. Extraordinary room-temperature photoluminescence in triangular WS$_2$ monolayers. Nano Lett 2012, 13, 3447-3454.

[124] Huang Y, Ding Z, Zhang W, et al. Gap States at Low-Angle Grain Boundaries in Monolayer Tungsten Diselenide. Nano Lett 2016, 16, 3682–3688.

[125] Lauritsen JV, Kibsgaard J, Helveg S, et al. Size-dependent structure of MoS$_2$ nanocrystals. Nat Nanotechnol 2007, 2, 53-58.

# Figures

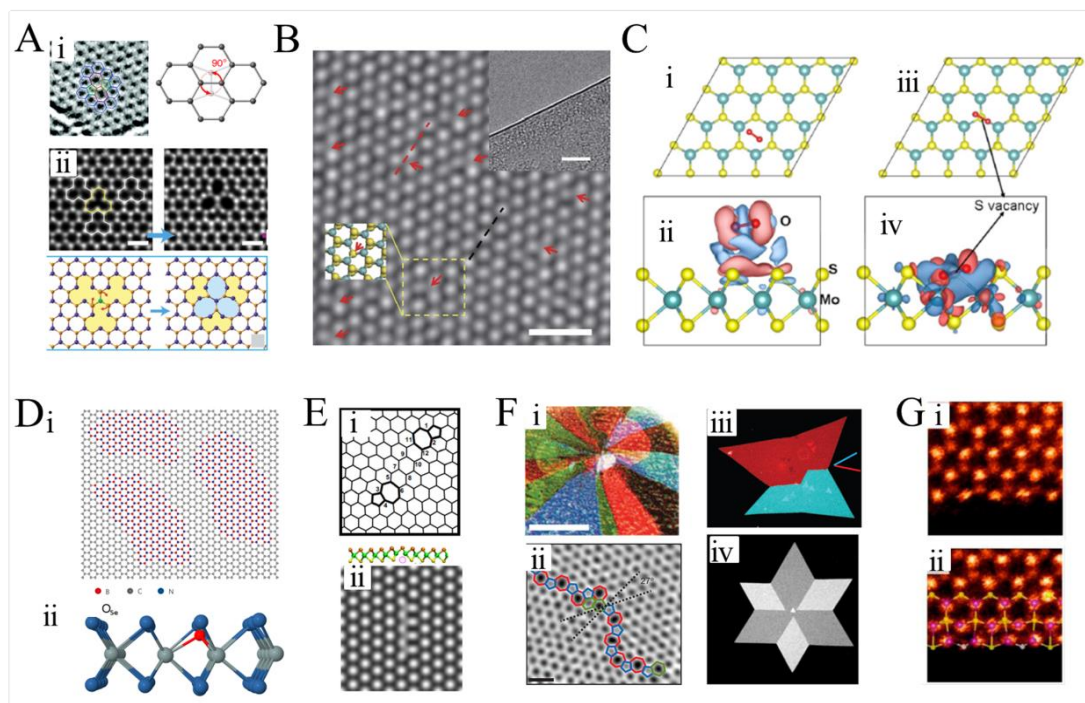

**Figure 1.** The types of defects in 2D materials. (A) SW defects. (i) HRTEM image of SW defects in graphene [51]. (ii) The filtered annular dark-field images of SW defects in WSe$_2$ (top) and the atomic model of T$_0$ to T$_1$ transformation (bottom) [34]. (B) Atomic structure of a monolayer MoS$_2$ by aberration-correct TEM [21]. The SVs are highlighted by red arrows. (C) Relaxed configuration and charge density difference of an O$_2$ molecule physisorbed on perfect monolayer MoS$_2$ (i, ii) and chemisorbed on defective monolayer MoS$_2$ (iii, iv) [33]. (D) Atomic model of substitutional impurities. (i). C atoms substituted by B and N atoms in graphen [62]. (ii) substitutional oxygen in TMDs [60]. (E) Line defects. (i) Line defects formed from aligned vacancy structures in graphene [65]. (ii) Single vacancy line defects in MoS$_2$ [66]. (F) GBs. (i, ii) TEM image of GBs in graphene [67]. (iii, iv) Dark-field TEM image of polycrystalline MoS$_2$ islands with tilt and mirror twin GBs [69]. (G) Edges [28]. (i) regular Mo-terminated edge of MoS$_2$. (ii) Reconstruction of the Mo-terminated edge.

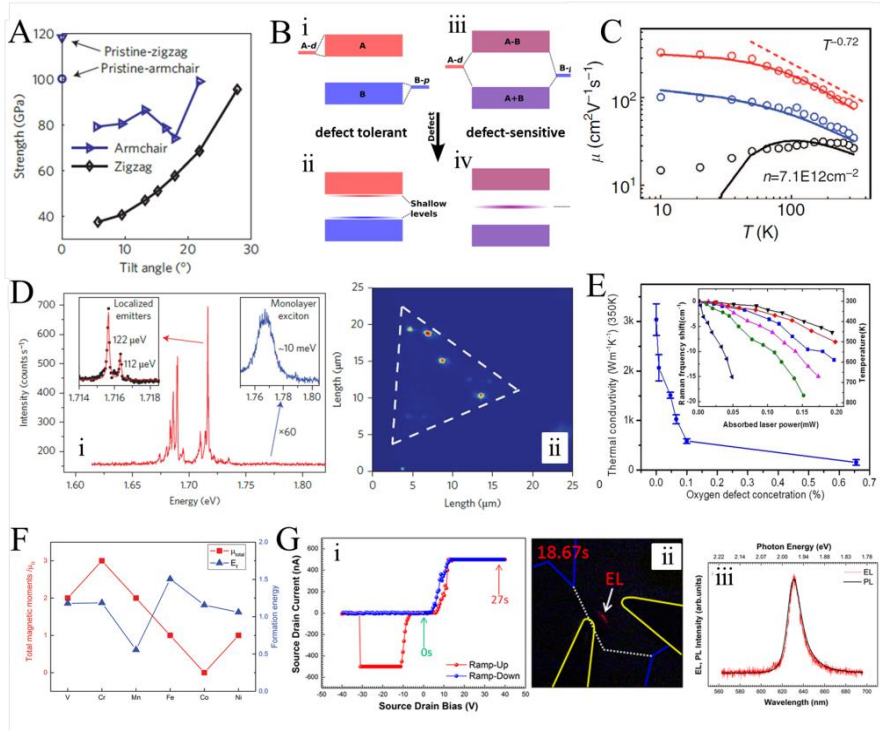

**Figure 2.** (A) The strength of GBs of graphene as a function of tilt angle [23]. (B) (i, iii) Band structures near the band edges for the defect-tolerant and defect-sensitive cases of TMDs, respectively. (ii, iv) Shallow and deep levels introduced after the creation of defects in TMDs [79]. (C) Mobility (μ) as a function of temperature (T) for as-exfoliated (black), top-side repaired (blue), double-side repaired (red) monolayer $MoS_2$ [80]. (D) (i) PL spectrum of localized emitters [86]. Inset is a high-resolution spectrum of highest intensity peak. (ii) PL intensity image of narrow emission lines within a spectral width of 12 meV centered at 1.719 eV. (E) Thermal conductivities of oxygen plasma treated graphene with different defect concentration. The inset is the G peak frequency shift as a function of the absorbed laser power [24]. (F) Calculated formation energies and total magnetic moments of BP doped with different transition metals elements [87]. (G) (i) A source drain bias ramping cycle used for EL generation (ii) EL images in the monolayer $WS_2$ GB region at forward bias. (iii) An EL spectrum is compared with a PL spectrum normalized to the EL intensity [90].

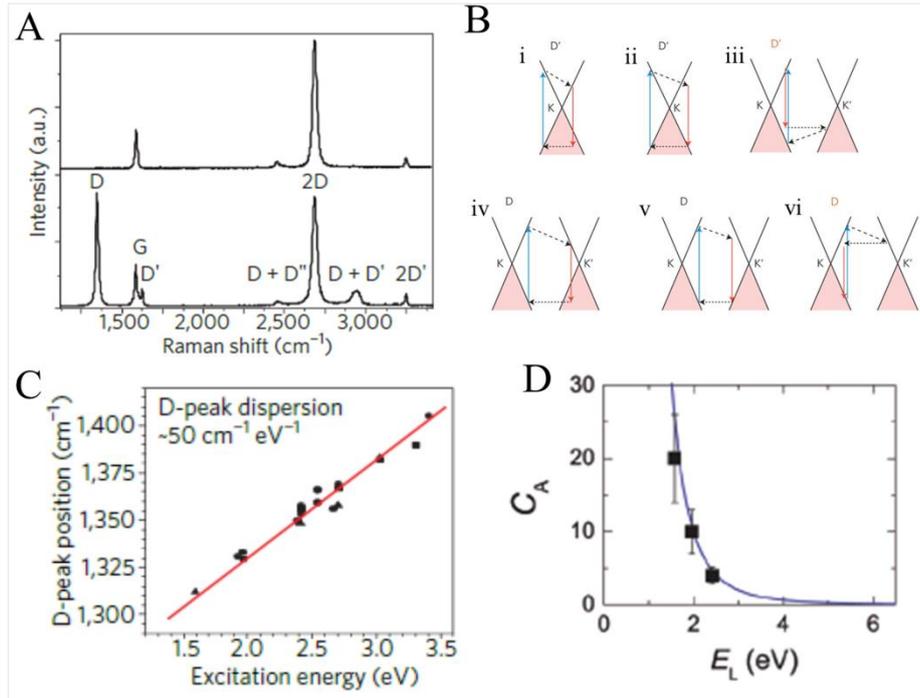

**Figure 3.** (A) Raman spectra of pristine (top) and defective (bottom) graphene. (B) Raman processes corresponding to D and D′ peaks of graphene. Electron dispersion (solid black lines), occupied states (shaded areas), interband transitions neglecting the photon momentum, accompanied by photon absorption (blue arrows) and emission (red arrows), intraband transitions accompanied by phonon emission (dashed arrows), electron scattering on a defect (horizontal dotted arrows). (C) Measured and calculated frequencies of the D peak as a function of the excitation energy [45]. (D) Laser energy dependence of the integrated Raman intensity ratio $I_D/I_G$, where $I_D/I_G \propto C_A$. The solid line is given by $C_A \propto E_L^{-4}$ fit [95].

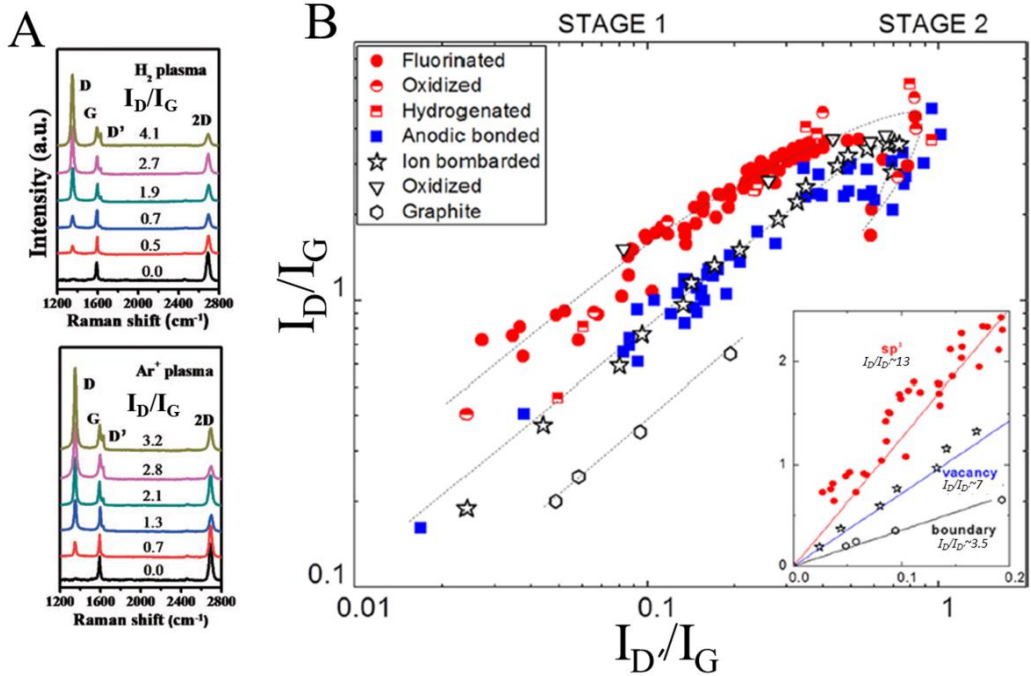

**Figure 4.** (A) Evolution of Raman spectra of (top) hydrogen- and (bottom) Ar+-plasma treated graphene [97]. (B) $I_D/I_G$ versus $I_D/I_G$ of graphene with different types of defects. The inset shows the linear dependence between the two parameters at low defect concentration, suggesting that graphene samples with different types of defects have different $I_D/I_D'$ [100].

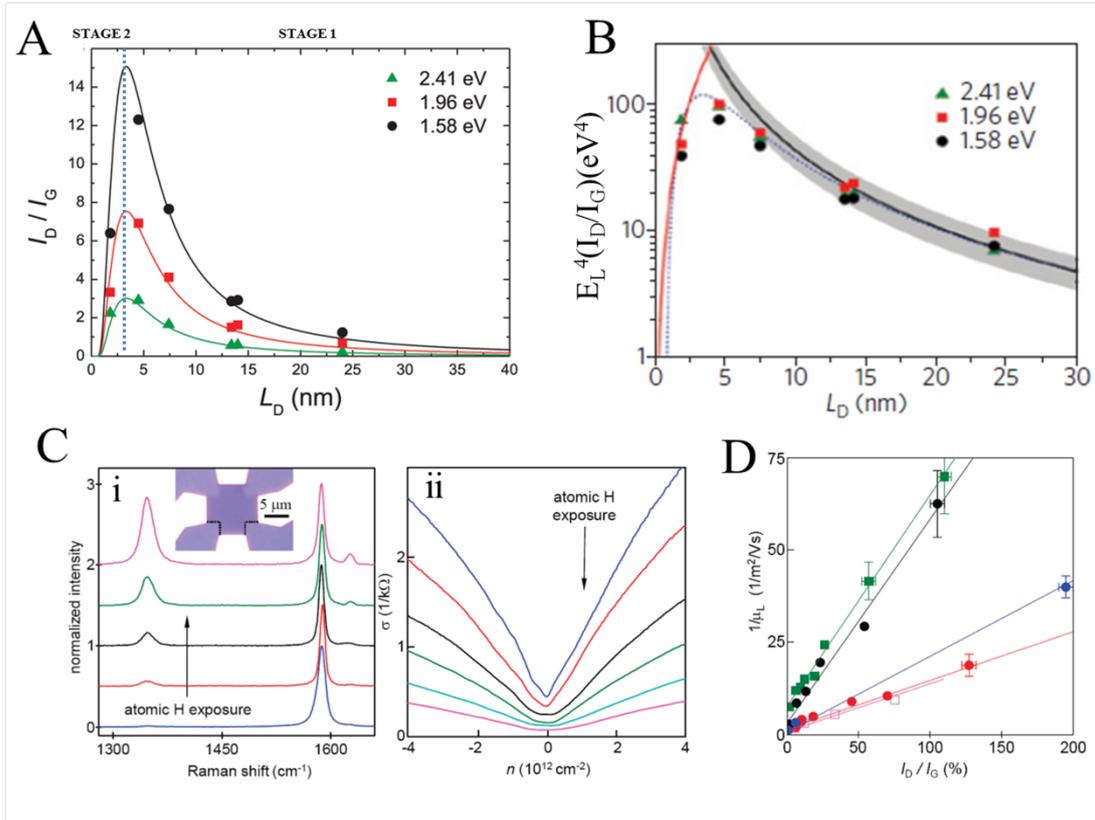

**Figure 5.** (A) The evolution of $I_D/I_G$ with interdefect distances $L_D$ under different laser energies [95]. (B) Amorphisation trajectory $E_L^4 [I_D/I_G]$ as a function of $L_D$ for different excitation energies [45]. (C) Hydrogen adatoms lead to an increase in the D peak intensity (i) and simultaneously affect graphene's electrical performance (ii). (D) Changes in mobility as a function of the D peak intensity. Different symbols denote different devices. Solid lines are the linear fits [14].

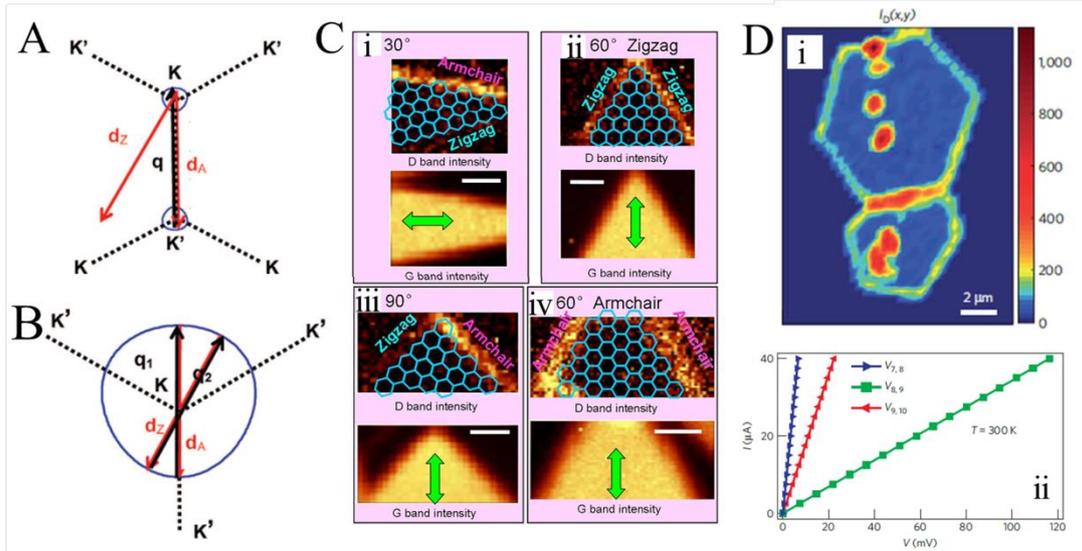

**Figure 6.** (A) Schematic of intervalley process: only the exchanged momentum from armchair edges $d_A$ can connect K and K′. (B) Schematic of intravalley process: both the exchanged momentum from armchair and zigzag edges, $d_Z$ and $d_A$, can connect points belonging to the same region around K (K′) [104]. The blue line is an iso-energy contour, taking into account the trigonal warping. (C) Raman images from graphene edges with angles (i) 30°, (ii) 60° (zigzag), (iii) 90° and (iv) 60° (armchair) [70]. (D) (i) D peak intensity images for two coalesced graphene grains with a single GB. (ii) Representative room-temperature I-V curves measured within each graphene grain ($V_{7,8}$, $V_{9,10}$) and across the grain boundary ($V_{8,9}$) [109].

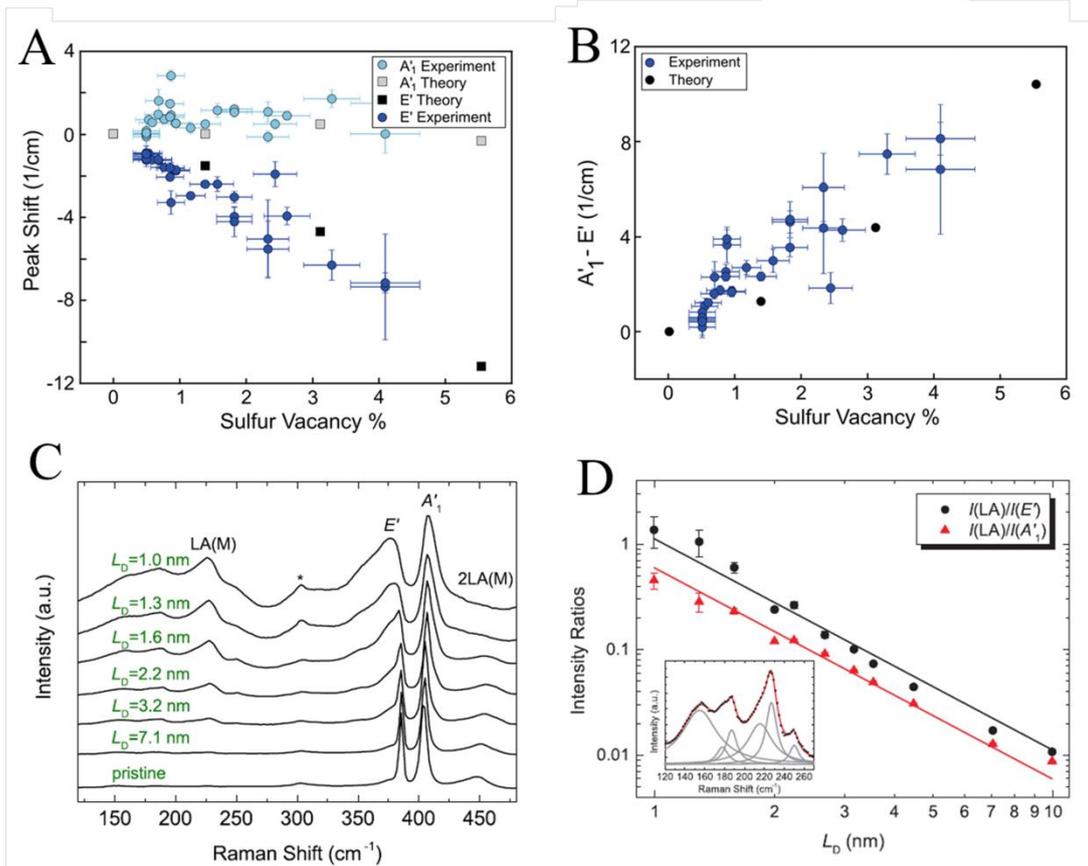

**Figure 7.** (A) Experimental and theoretical peak shifts of the E′ and A′$_1$ peaks as a function of sulfur vacancy percentage. (B) Change in frequency difference between the E′ and A′$_1$ peaks as a function of defect concentration [111]. (C) Raman spectra of MoS$_2$ flakes with varying interdefect distances L$_D$. (D) Experimental intensity ratios I(LA)/I(A′$_1$) and I(LA)/I(E′) of MoS$_2$ as a function of inter-defect distance L$_D$. The inset shows the low frequency peaks and the corresponding Lorentzian fits, where the LA peak is located at ~227 cm$^{-1}$ [47].

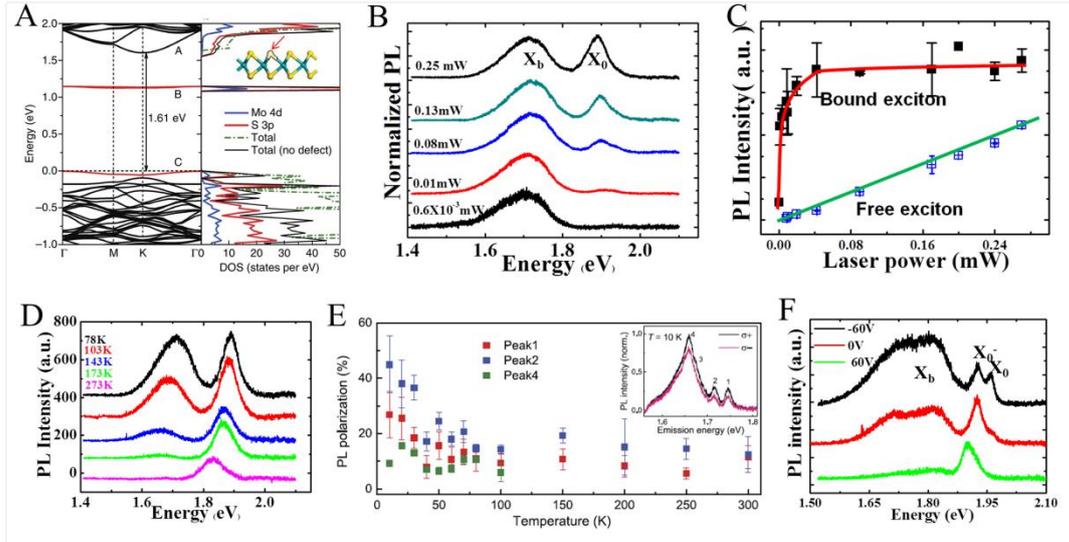

**Figure 8.** (A) Band structure (left) and partial density of states (right) of a monolayer MoS$_2$ 5×5 supercell with an SV [21]. The localized states are highlighted by red lines. Green dashed line corresponds to the case without SV. (B) Laser power dependence of PL spectra of MoS$_2$, normalized by the intensity of X$_b$. (C) PL intensity of bound exciton and free exciton of MoS$_2$ with increasing laser power. (D) Temperature dependence of PL spectra for MoS$_2$. (E) Temperature dependence of the PL circular polarization of monolayer WSe$_2$. Inset: Polarization-resolved PL spectra for σ$^+$ and σ$^−$ detections [116]. (F) Gate dependence of PL spectra for MoS$_2$ measured at 83 K.

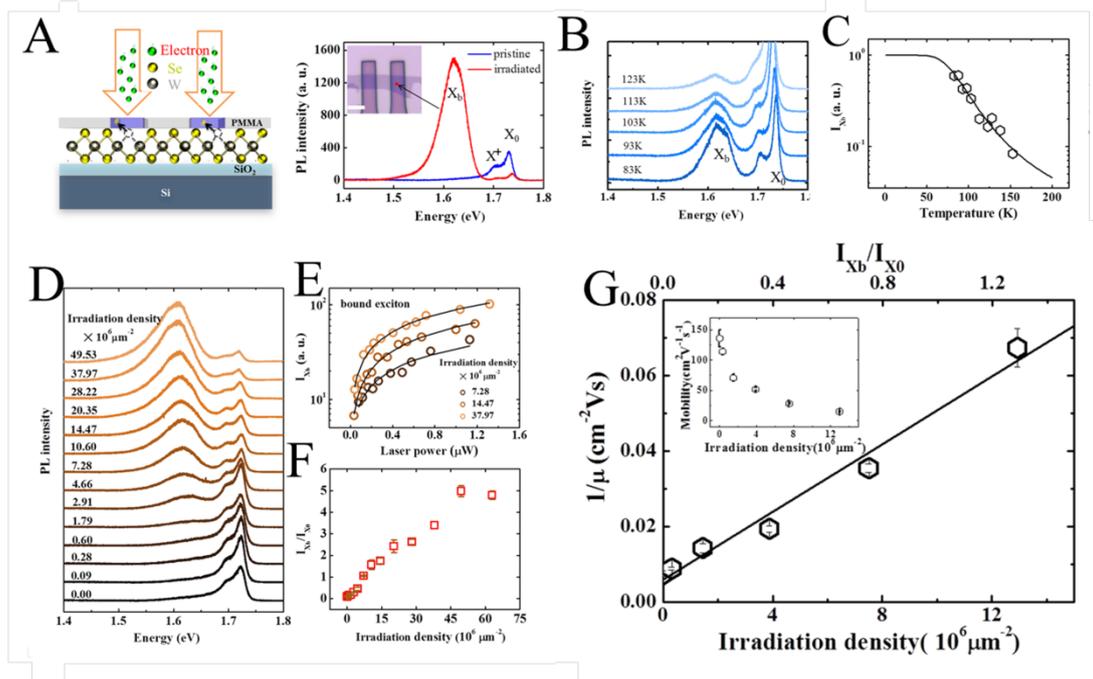

**Figure 9.** (A) (Left) Schematic diagram of electron-beam irradiation on monolayer WSe$_2$ sample during the EBL process. (right) PL spectrum of monolayer WSe$_2$ with and without electron-beam irradiation. The inset shows optical image of WSe2 with PMMA patterned by EBL, scale bar is 5 μm. Temperature dependence of PL spectra (B) and X$_b$ intensity (C) of WSe$_2$ after electron-beam irradiation. (D) PL spectra of monolayer WSe$_2$ under different irradiation density. (E) Laser power dependence of the intensity of X$_b$ under different irradiation density. (F) The dependence of I$_{Xb}$/I$_{X0}$ on irradiation density. (G) Changes of scattering rate 1/μ and mobility μ (inset) as a function of electron-beam dosage and I$_{Xb}$/I$_{X0}$ [48].

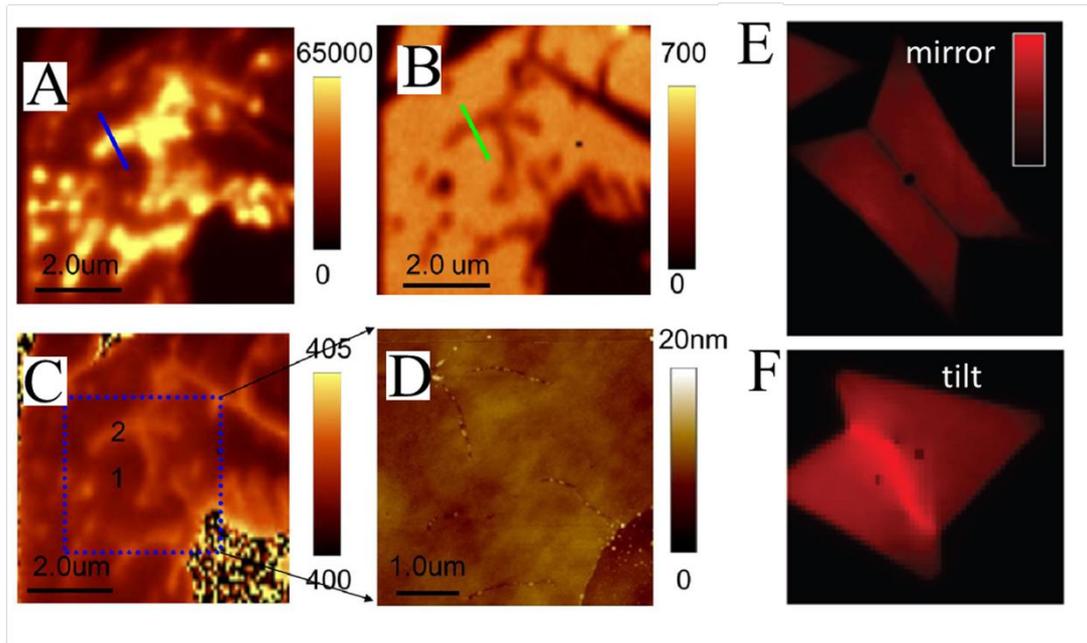

**Figure 10.** PL intensity (A), Raman $A'_1$ peak intensity (B), and $A'_1$ peak frequency (C) images of a monolayer $MoS_2$ after annealing for 1 h at 500 °C at 0.1 Pa. (D) AFM image corresponding to the blue square in (C) [33]. PL intensity images of CVD $MoS_2$ islands with mirror (E) and tilt (F) GBs. The mirror twin boundary has a 50% quenching of PL, while the tilt boundary has a 100% enhancement of PL [69].